\documentclass[prb,twocolumn,showkeys,showpacs,preprintnumbers,amsmath,amssymb,floatfix,superscriptaddress]{revtex4}

\usepackage{graphicx}
\usepackage{dcolumn}
\usepackage{bm}
\usepackage{dsfont}
\usepackage{float}
\usepackage[section]{placeins}

\begin{document}

\title{Pulsed Low-Field Electrically Detected Magnetic Resonance}
\author{L. Dreher}
\email{dreher@wsi.tum.de} 
\affiliation{Walter Schottky Institut, Technische Universit\"at M\"unchen, Am Coulombwall 4, 85748 Garching, Germany}
\author{F. Hoehne}
\affiliation{Walter Schottky Institut, Technische Universit\"at M\"unchen, Am Coulombwall 4, 85748 Garching, Germany}
\author{H. Morishita}
\affiliation{Keio University, 3-14-1 Hiyoshi Kouhoku-Ku, Yokohama 223-8523, Japan}
\author{H. Huebl}
\affiliation{Walther-Mei\ss ner-Institut, Bayerische Akademie der Wissenschaften, Walther-Mei\ss ner-Strasse 8, 85748 Garching, Germany}
\author{M. Stutzmann}
\affiliation{Walter Schottky Institut, Technische Universit\"at M\"unchen, Am Coulombwall 4, 85748 Garching, Germany}
\author{K. M. Itoh}
\affiliation{Keio University, 3-14-1 Hiyoshi Kouhoku-Ku, Yokohama 223-8523, Japan}
\author{M. S. Brandt}
\affiliation{Walter Schottky Institut, Technische Universit\"at M\"unchen, Am Coulombwall 4, 85748 Garching, Germany}

\date{\today}

\begin{abstract}
We present pulsed electrically detected magnetic resonance (EDMR) measurements at low magnetic fields using phosphorus-doped silicon with natural isotope composition as a model system. Our measurements show that pulsed EDMR experiments, well established at $X$-band frequencies (10~GHz), such as coherent spin rotations, Hahn echoes, and measurements of parallel and antiparallel spin pair life times are also feasible at frequencies in the MHz regime. We find that the Rabi frequency of the coupled $^{31}$P electron-nuclear spin system scales with the magnetic field as predicted by the spin Hamiltonian, while the measured spin coherence and recombination times do not strongly depend on the magnetic field in the region investigated. The usefulness of pulsed low-field EDMR for measurements of small hyperfine interactions is demonstrated by electron spin echo envelope modulation measurements of the P$_\text{b0}$ dangling-bond state at the Si/SiO$_2$ interface. A pronounced modulation with a frequency at the free Larmor frequency of hydrogen nuclei was observed for radio frequencies between 38~MHz and 400~MHz, attributed to the nuclear magnetic resonance of hydrogen in an adsorbed layer of water. This demonstrates the high sensitivity of low-field EDMR also for spins not directly participating in the spin-dependent transport investigated.
\end{abstract}



\maketitle
\section{Introduction}

Electron paramagnetic resonance (EPR) is a versatile spectroscopic tool for the investigation of the microscopic structure and local environment of paramagnetic centers in molecules and in solids.\cite{AC103_223} Elaborate EPR techniques such as electron nuclear double resonance (ENDOR)\cite{PR114_1219,PRSA283_452,PLA47_1,AC103_223} and electron spin echo envelope modulation (ESEEM)\cite{AC103_223,rowan_electron-spin-echo_1965,mims_envelope_1972} allow for a characterization of isotropic and anisotropic hyperfine interactions of the center's electron spin with nuclear spins in its vicinity, allowing, e.g., for the mapping of the electronic wave function and are thus a sensitive probe to obtain information on the local environment.\cite{AC103_223} Besides its application in materials science, pulsed EPR has recently attracted much attention in the context of quantum computing proposals where quantum bits are realized by electron or nuclear spins which can be manipulated using pulsed EPR and ENDOR.\cite{N414_883,N455_1085}

For both applications, spectroscopy and quantum computation, it can be advantageous to employ rather small magnetic fields, where the electron Zeeman interaction is comparable with the hyperfine interaction.\cite{PRL105_67602,JAP109_102411,S336_1280} At these fields, the product states of the electron and nuclear spin system become mixed, thereby shortening the time it takes to manipulate the nuclear spin. This is referred to as hyperfine enhancement.\cite{HighResEPR} Particularly in systems with rather small hyperfine interactions, it would be beneficial to employ low magnetic fields especially for ENDOR and ESEEM measurements to enable the investigation of these small hyperfine interactions. For applications in quantum computation, this magnetic field region is of interest because a hybrid qubit consisting of a mixed electron- and nuclear-spin state can be realized and the extent to which the qubit is electron- or nuclear-spin like can be tuned by the external field. This has in particular been investigated for Si:Bi, where, due to the large hyperfine coupling of Bi, the mixing of electron and
nuclear spin occurs at relatively large magnetic fields, allowing for conventional $X$-band EPR.\cite{mohammady_analysis_2012-1,wolfowicz_atomic_2013,morley_quantum_2013} 
Conventional EPR however, requires a thermal equilibrium polarization of the spin system, limiting the sensitivity of conventional EPR to about $10^{9}$ spins at $X$-band frequencies and typical cryogenic temperatures of the order of 5~K, making low-field experiments challenging. However, by using optical and electrical spin-readout schemes, known as optically/electrically detected magnetic resonance (ODMR/EDMR), single electron and nuclear spins can be detected, their spin state determined,\cite{PRL92_76401,morello_single-shot_2010,pla_high-fidelity_2013} and the toolbox of sophisticated pulse sequences realized in EPR can be adapted to pulsed EDMR and ODMR. Rather than relying on a thermal polarization of the spin system, these readout mechanisms are based on spin selection rules, which influence the electrical transport through a sample or device, or change the photoluminescence of paramagnetic states. This makes these detection schemes sensitive even at comparatively low magnetic fields where the Zeeman splitting of the spin states is small compared to the thermal energy. In phosphorus-doped silicon, continuous-wave (cw) EDMR has been demonstrated at magnetic fields of below 1~mT,\cite{PRB80_205206} where the Zeeman splitting of the electron spins is of the order of tens of MHz. Furthermore, low-field cw EDMR has recently been employed to demonstrate the wide magnetic field range of a potential magnetometer based on thin-film organic devices.\cite{NC3_898} To our knowledge, however, pulsed EDMR experiments have not been reported in the low magnetic field region, whereas at $X$-band frequencies and the corresponding magnetic fields coherent electron and nuclear spin manipulations, \cite{NP2_835,PRL106_187601,harneit_room_2007,mccamey_spin_2008,PRB81_75214} electron and nuclear spin echo measurements,\cite{PRL100_177602,PRL108_27602} electron double resonance experiments,\cite{PRL104_46402,suckert_electrically_2013} and electron spin echo envelope (ESEEM) measurements,\cite{PRL106_196101,fehr_electrical_2011} have been demonstrated in the past years. Only at exactly zero magnetic field and for a spin 1 center with large zero-field splitting, pulsed EDMR experiments have been reported.\cite{franke_spin-dependent_2014}

Using phosphorus-doped silicon as a model system, we demonstrate in this work, that the toolbox of pulsed EDMR, established at $X$-band and higher\cite{meier_multi-frequency_2013,morley_long-lived_2008} frequencies, can be transferred to the radio frequency (rf) regime. In doing so, the main objective of this work is to show the usefulness of pulsed low-field EDMR for spectroscopic applications, in particular for measurements involving small hyperfine interactions. Our experimental results show that using low-field electrically detected ESEEM (EDESEEM) small hyperfine couplings that are not detectable at $X$-band frequencies can be readily observed, rendering pulsed low-field EDMR a very promising tool for interface science.

 This manuscript is structured as follows. The experimental details are introduced in Sec.~\ref{sec:exp_details}. We start in Sec.~\ref{sec:results_and_discussion} with an introduction to the spin Hamiltonian of the $^{31}$P donor reviewing the level-mixing of electron and nuclear spins at low magnetic fields and in Sec.~\ref{sec:Rabis} we present measurements of coherent spin oscillations of the coupled electron-nuclear spin system. We then turn to the investigation of the spin coherence times (Sec.~\ref{sec:HahnEcho}) and the measurement of the antiparallel (Sec.~\ref{sec:InvRec}) and parallel spin-pair recombination times (Sec.~\ref{sec:DPS}). In Sec.~\ref{sec:ESEEM_Pb}, we show stimulated echo decay measurements of the P$_\text{b0}$ center at low magnetic fields, which exhibit a pronounced ESEEM effect that is not observed at $X$-band frequencies. Finally, we summarize our findings and give an outlook to further experiments (Sec.~\ref{sec:Summary_Outlook}). 

\section{Experimental Details}\label{sec:exp_details}
The EDMR mechanism employed is based on a spin-dependent recombination process involving a spin-pair consisting of the $^{31}$P electron spin (with $S=1/2$) and a  dangling-bond defect (also with $S=1/2$) at the Si/SiO$_2$ interface, referred to as P$_\text{b0}$.\cite{JAP83_2449} In the absence of above band-gap illumination, the $^{31}$P donors are compensated by the dangling bond states, thus the $^{31}$P close to the interface are ionized ($^{31}$P$^+$) and the P$_\text{b0}$ are negatively charged (P$_\text{b0}^-$). If the sample is illuminated with above-bandgap light, the $^{31}$P$^+$ captures an electron from the conduction band and one P$_\text{b0}^-$ electron can undergo a transition into an empty valence band state. The resulting neutral donor and the neutral dangling bond form the weakly coupled $^{31}$P-P$_\text{b0}$ spin pair, with a coupling constant of $\approx 1$~MHz.\cite{PRL104_46402,suckert_electrically_2013} Owing to the Pauli principle, the spin pair with parallel spin orientation is long lived,\cite{JdP39_51} with a lifetime $\tau_\mathrm{p}$ of the order of one millisecond at 5 K, while the antiparallel spin pairs recombine on a time scale $\tau_\mathrm{ap}$ of the order of 10~$\mu$s.\cite{PRL108_27602,hoehne_time_2013} Thus, a steady state is established, in which in good approximation only spin-pairs with parallel spin orientation exist. If either one of the electron spins is flipped by a resonant microwave pulse, the spin pair is transformed into the antiparallel configuration, leading to recombination and thus a quenching of the photocurrent which is monitored in the experiment.

The sample studied in this work consists of a 20~nm-thick phosphorus-doped ([P]=$3\times 10^{16}$cm$^{-3}$) Si layer with natural isotopic composition grown on a nominally undoped 2.5~$\mu$m-thick Si buffer layer on top of a [001]-oriented silicon-on-insulator substrate; the P-doped top layer is covered by a native oxide. To measure the photocurrent through the sample, Cr/Au interdigital contacts with a period of 20~$\mu$m covering an active area of 2$\times$2.5~mm$^2$ were applied using optical lithography. The sample was placed in an external magnetic field oriented along the [110] crystal axis within a dielectric microwave resonator equipped with radio frequency (rf) coils designed for pulsed ENDOR. It was cooled down to 5~K and illuminated with the light of a pulsed LED (Thorlabs LDC 210 controller) with a rise time of $\approx2$~$\mu$s and a wavelength of 625~nm. A bias voltage of 300~mV was symmetrically applied using a transimpedance amplifier with built-in high- and low-pass filters at cut-off frequencies of 2~kHz and 200~kHz, respectively. The light intensity was adjusted such that the photocurrent was 50~$\mu$A for all experiments under cw illumination. To achieve sufficiently short rf $\pi$-pulse lengths, a 300~W rf amplifier was connected to a capacitive matching network, consisting of a variable input capacitance and a variable capacitance to ground, which was attached to one end of the rf coil. A 50~$\Omega$ power load was connected to the other connector of the coil and used to monitor the rf pulses by an oscilloscope. The capacitors were adjusted such that the reflected power at the input of the circuitry is minimal at the desired frequency, resulting in a bandwidth of the matching network of typically 3-10~MHz. The achieved $\pi$-pulse lengths were in the range of  50-80~ns for unmixed electron-spin transitions. To minimize the coupling of the high-power rf-pulses into the photocurrent circuitry, the sample was carefully aligned within the resonator such that the active contact area was centered in between the pair of rf-coils. Additionally, in-house-built low-pass filters at cut-off frequencies of 10~MHz were used to protect the input of the transimpedance amplifier. For experiments involving a spin echo, we employed a lock-in technique realized by phase cycling the final projection pulse of the spin echo by 180 degrees from shot to shot, subtracting the subsequent current transients from each other, and averaging the integrated current transient over 200-500 cycles by a fast digitizing board (Gage Applied) as described in detail in Ref.~\onlinecite{RoSI83_43907}, yielding a charge $\Delta Q$ as the primary observable of the experiment. This detection scheme suppresses low frequency noise and reduces non-resonant contributions to the current transient. In contrast to previous experiments,\cite{PRB83_235201,PRL106_187601,PRL106_196101,PRL108_27602,RoSI83_43907} the small bandwidth of the matching network results in a comparatively long ring-down time of the rf pulses, leading to an overlap of pulses for short inter-pulse delays. This overlap causes non-resonant contributions to the photocurrent transient which are not entirely removed by phase cycling only the final projection pulse of the echo. Therefore, we extended the two-step phase cycle to a four step phase cycle,\cite{Schweiger_Book} additionally switching the phase of the first rf pulse by 180$^\circ$ to further reduce non-resonant signals. We thus recorded for each data point two datasets as described above and in Ref.~\onlinecite{RoSI83_43907}, alternating the phase of the first rf pulse in the spin echo by 180 degrees. By subtracting these data sets from each other, the non-resonant contributions to the signal could be further reduced.

\section{Results and Discussion}\label{sec:results_and_discussion}
In this section, we will present and discuss the pulsed low-field experiments performed. Figure~\ref{fig:OverviewLF} (a) shows the Breit-Rabi diagram of the $^{31}$P donor in the magnetic field region of interest in this work. To describe this spin system consisting of an electron spin ($S=1/2$) and a nuclear spin ($I=1/2$) coupled by an isotropic hyperfine interaction ($A$=117.5~MHz),\cite{PR114_1219} it is sufficient to consider the spin Hamiltonian
\begin{equation}
\hat{H}=g\mu_\text{B}B \hat{S}_z+ A \bf{\hat{S}}\cdot \bf{\hat{I}},
\label{eq:SpinHamiltonian}
\end{equation}
where $g=1.9985$ is the $g$-factor of the phosphorus electron spin,\cite{PR114_1219} $\mu_\text{B}$ is the Bohr magneton in frequency units, $B$ is the magnitude of the external magnetic flux density, defining the quantization axis $z$, and $\bf{\hat{S}}$ and $\bf{\hat{I}}$ are the dimensionless electron and nuclear spin operators, respectively. Here, we have disregarded the nuclear Zeeman interaction, which is negligible compared with the hyperfine interaction and the electron Zeeman interaction. The energy eigenstates of Eq.~\eqref{eq:SpinHamiltonian} in the order of descending energy read as
\begin{eqnarray}
\left|1\right>&=&\left|\uparrow \Uparrow\right>\nonumber\\
\left|2\right>&=&\cos(\eta/2)\left|\uparrow \Downarrow\right>+\sin(\eta/2)\left|\downarrow \Uparrow\right>\nonumber\\
\left|3\right>&=&\left|\downarrow \Downarrow\right>\nonumber\\
\left|4\right>&=&\cos(\eta/2)\left|\downarrow \Uparrow\right>-\sin(\eta/2)\left|\uparrow \Downarrow\right>\label{eq:Eigenstates},
\end{eqnarray}
where $\uparrow$($\downarrow$) and $\Uparrow$($\Downarrow$) denote electron and nuclear spins with $\hat{S}_z$ or $\hat{I}_z$ eigenstates of 1/2 (-1/2), respectively. The mixing angle $\eta$ is given by
\begin{equation}
\tan (\eta) = A/(g\mu_\text{B}B)
\label{eq:mixingangle}
\end{equation}
and determines the extent to which the high-field states, defined by the kets on the right-hand side of Eqs.~\eqref{eq:Eigenstates}, are mixed.
A circularly polarized rf magnetic field with amplitude $B_1$ oriented perpendicularly to the static field $B$ with the rf in resonance with the frequency splitting of two eigenstates $\left|i\right>$ and $\left|j\right>$ will drive transitions between these states with an angular frequency given by
\begin{equation}
\omega_\text{Rabi}= \underbrace{\frac{g \mu_\text{B}}{\hbar}}_{\gamma} \left<i\left|\hat{S}_x\right|j\right>B_1 =\frac{1}{2}\underbrace{\gamma\left<i\left|2\hat{S}_x\right|j\right>}_{\gamma_\text{x,eff}}B_1,
\label{eq:Rabi_frequency}
\end{equation}
assuming $B\gg B_1$. In the following, we will refer to the transitions $\left|1\right>\leftrightarrow \left|4\right>$ and $\left|2\right>\leftrightarrow\left|3\right>$ as electron-spin-like transitions (P$_\text{e}$) since these transitions are the allowed transitions according to the dipole selection rules in the high-field limit ($\Delta m_S\pm1$, where $m_S$ are the eigenvalues of $\hat{S}_z$). The corresponding matrix element in Eq.~\eqref{eq:Rabi_frequency} is given by $\gamma_\text{x,eff}/\gamma=\cos(\eta/2)$, which asymptotically approaches unity in the limit of large magnetic fields as shown by the solid lines in Fig.~\ref{fig:OverviewLF} (c). Accordingly, the transitions $\left|1\right>\leftrightarrow\left|2\right>$ and $\left|3\right>\leftrightarrow\left|4\right>$ are referred to as nuclear-spin like transitions (P$_\text{n}$) since these transitions are allowed according to the dipole selection rules for nuclear spins in the high-field limit ($\Delta m_I\pm1$, where $m_I$ are the the eigenvalues of $\hat{I}_z$). The effective gyromagnetic ratio for this transition, neglecting the coupling of the driving field to the nuclear spin, reads as $\gamma_\text{x,eff}/\gamma=\sin(\eta/2)$ and asymptotically vanishes as $B$ increases. We note that at a magnetic field of 0.35~T typically used for ESR experiments at $X$-band frequencies, $\sin(\eta/2)=0.006$, resulting in $\sin(\eta/2)\gamma/\gamma_\text{n}\approx 10$ with the nuclear $^{31}$P gyromagnetic ratio $\gamma_\text{n}$.\cite{PR114_1219} Thus, in an ENDOR experiment at $X$-band frequencies the coupling of the electron spin to the rf driving field for this forbidden transition is, due to the hyperfine interaction, still 10 times larger than that of the nucleus for this allowed nuclear spin transition, leading to shorter $\pi$-pulse times. Assuming a system with a rather weak hyperfine interaction of 1 MHz, a magnetic resonance experiment has to be performed at roughly 100~MHz to achieve the same hyperfine enhancement obtained for Si:P at 10~GHz ($X$-band), demonstrating the advantage of low-field EDMR for systems with small hyperfine interactions.
\subsection{Rabi Oscillations of Mixed Electron-Nuclear Spin Transitions}\label{sec:Rabis}

\begin{figure*}[!H]
\includegraphics[]{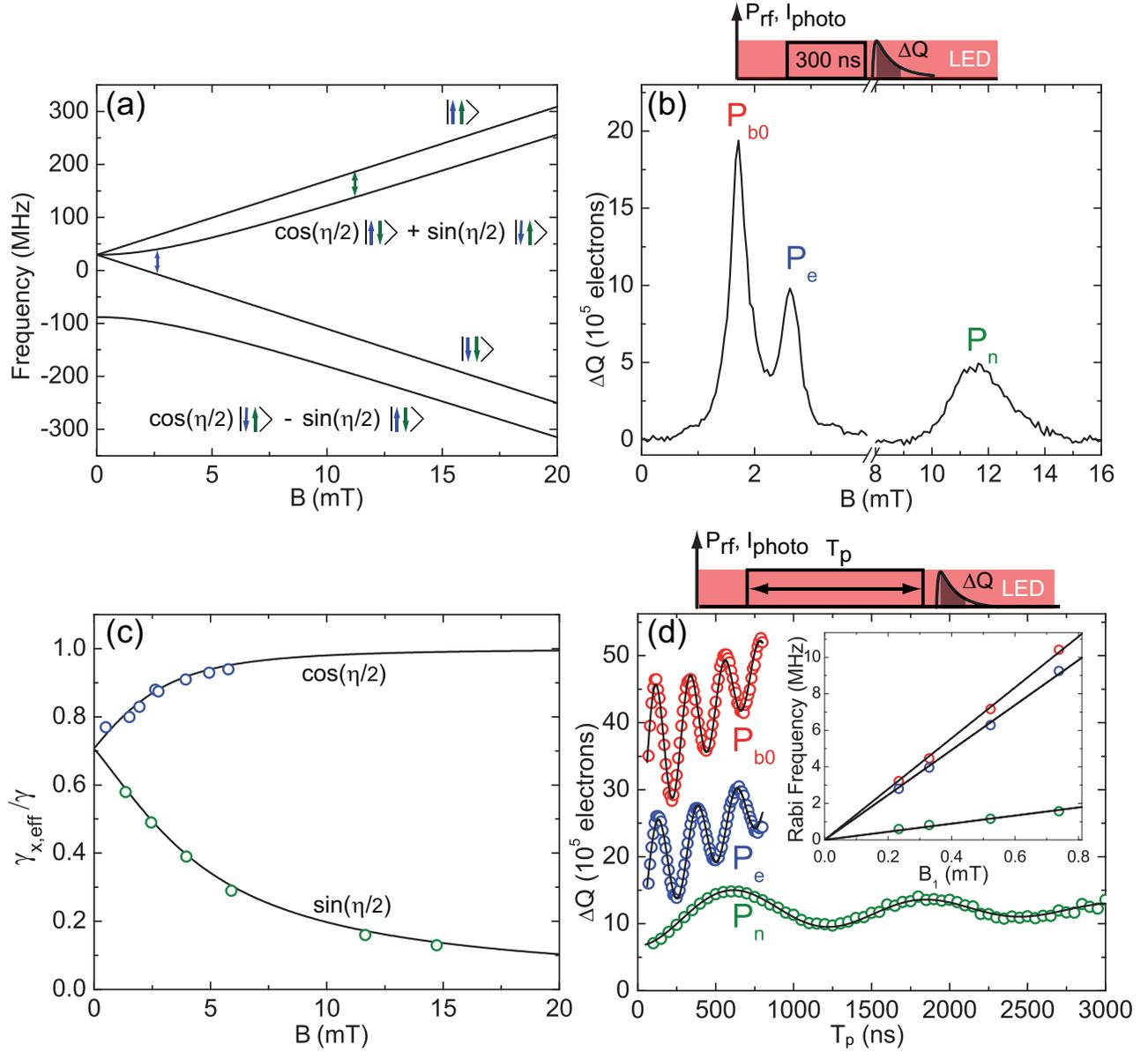}
\caption{(a) Breit-Rabi diagram of the $^{31}$P donor. Due to the level mixing, electron- and nuclear-spin like transitions become observable in a pulsed low-field EDMR experiment as shown in (b). The spin system was excited with a 300~ns-long square rf pulse with a frequency of 48.6~MHz. The resonance line of the nuclear-spin like transition is broader than the one of the electron-spin like transition because the magnetic field dependence of the transition frequency is weak as shown in (a). (d) Rabi oscillations measured on the electron-spin like, the nuclear-spin like, and the dangling-bond transition (exemplarily shown for an rf of 48.6~MHz). The solid black lines represent fits with exponentially damped cosines and a linear background accounting for the reduced spectral width of the longer pulses;\cite{PRL104_46402} the data are vertically offset for better visibility. Further Rabi oscillations were measured as a function of the rf power level, resulting in a linear dependence of the Rabi frequencies ($\omega_\text{Rabi}/2\pi$) on the rf magnetic field $B_1$ as shown in the inset; the P$_\text{b0}$ transition was used for the calibration of the $B_1$ field and the experimental values for $\gamma_\text{x,eff}/\gamma$ were extracted from the slopes of the linear $B_1$-Rabi-frequency relationship of the electron- and nuclear-spin like transitions.  (c) The effective gyromagnetic ratio $\gamma_\text{x,eff}/\gamma$, which determines the coupling of a spin transition to the rf driving field, is given by $\cos(\eta/2)$ and $\sin(\eta/2)$ for electron- and nuclear-spin like transitions, respectively, and plotted against the magnetic field (solid lines). The blue and green data points shown were extracted from the $B_1$ dependence of the Rabi frequencies of the electron- and nuclear-spin like transitions, cf.~the inset in (d).}
\label{fig:OverviewLF}
\end{figure*}

To experimentally verify Eq.~\eqref{eq:Rabi_frequency}, we measured coherent spin oscillations employing different radio frequencies such that the resulting $\gamma_\text{x,eff}/\gamma$ covers the range from 0.1 to 0.9. To this end, we started at each rf by measuring a pulsed magnetic field spectrum as exemplarily shown for an rf of 48.6~MHz in Fig.~\ref{fig:OverviewLF} (b). The spectrum shown was recorded by applying a 300~ns-long rf pulse under cw illumination and by recording and boxcar averaging the current transient immediately after the pulse for each magnetic field point, providing the charge $\Delta Q$ as signal intensity. In this  paper, the pulse sequences are sketched on top of the results shown. We observe three resonances: one of them stems predominantly from the P$_\text{b0}$ center with contributions from other weaker signals at $g\approx 2$ which are not spectrally resolved at the frequencies employed in this work (cf.~e.g.~Ref.~\onlinecite{PRB83_235201} for an example of a spectrum of a similar sample at $X$-band frequency where the signals around $g\approx 2$ are spectrally resolved). The other two signals, labeled P$_\text{e}$ and P$_\text{n}$, correspond to the electron-spin like and nuclear-spin like transitions sketched by blue and green arrows in Fig.~\ref{fig:OverviewLF} (a), respectively. The P$_\text{n}$ peak is broader compared to the P$_\text{e}$ peak, reflecting the weaker field-dependence of the P$_\text{n}$ transition frequency compared to the P$_\text{e}$ transition frequency visible in Fig.~\ref{fig:OverviewLF} (a). 

To observe coherent spin oscillations, we measured the integrated photocurrent response to rf pulses by varying the pulselengths at each of the three resonance fields and at two additional off-resonant field points to subtract non-resonant contributions to the current transients. The resulting Rabi oscillations recorded at 48.6~MHz are shown in Fig.~\ref{fig:OverviewLF} (d), already demonstrating that the coupling to the driving field is weakest for the P$_\text{n}$ transition, manifesting itself in the rather slow oscillation. The Rabi frequency for the P$_\text{e}$ transition is still substantially slower than that of the purely electronic P$_\text{b0}$ transition. To quantify the coupling, we measured Rabi oscillations at different rf power levels with the results shown in the inset in Fig.~\ref{fig:OverviewLF} (d). Since the P$_\text{b0}$ Rabi frequency should be independent of the static $B$ field, the P$_\text{b0}$ transition was used to calculate the rf $B_1$ field at a given power level for each rf employed. From a linear fit of the $B_1$ dependencies of the Rabi frequencies [cf.~inset in Fig.~\ref{fig:OverviewLF} (d)], we extract $\gamma_\text{x,eff}/\gamma$ for the P$_\text{e}$ and P$_\text{n}$ transitions, and plot the result in Fig.~\ref{fig:OverviewLF} (c) together with $\cos(\eta/2)$ and $\sin(\eta/2)$ as a function of $B$, demonstrating an excellent agreement of the experiment with Eq.~\eqref{eq:Rabi_frequency}.

\subsection{Hahn Echo Measurements: Coherence Times}\label{sec:HahnEcho}

\begin{figure*}[!htp]
\includegraphics[]{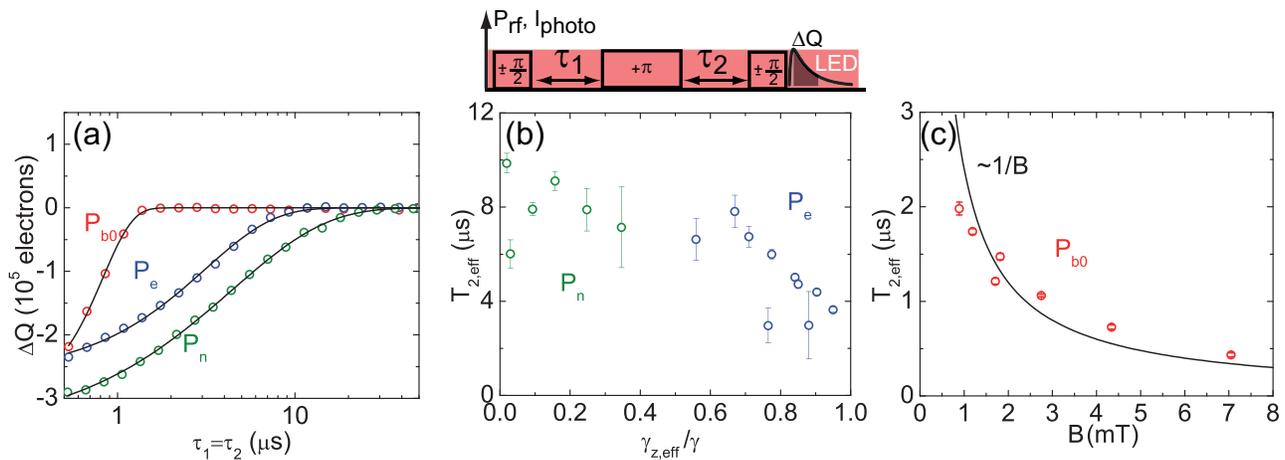}
\caption{On top: Pulse sequence employed for low-field spin echo measurements. (a) Spin echo decays of the P$_\text{b0}$, P$_\text{e}$, and P$_\text{n}$ transitions at an rf of 50.5~MHz with stretched exponential fits. (b) Effective coherence times of the P$_\text{e}$ and P$_\text{n}$ transitions as a function of $\gamma_{z,\text{eff}}/\gamma$. (c) Effective coherence time of the P$_\text{b0}$ transition as a function of the magnetic field.}
\label{fig:EchoDecay}
\end{figure*}

Having verified the mixing-angle dependence of the Rabi frequencies, we turn to the investigation of the spin coherence times. To this end, we performed spin echo decay measurements\cite{PRL100_177602} using the pulse sequence depicted in Fig.~\ref{fig:EchoDecay}. The $\pm$ signs in the pluse schemes indicate that the phase of the respective rf was cycled by 180$^\circ$ as discussed in Sec.~\ref{sec:exp_details}. The $\pi$-pulse length for each transition was extracted from the Rabi measurements and was typically 50-80~ns for the electron-spin like transitions. Figure~\ref{fig:EchoDecay} (a) exemplarily shows the measured echo decays for the P$_\text{b0}$, the P$_\text{e}$, and the P$_\text{n}$ transitions at an rf of 50.5~MHz with stretched exponential [$\exp((\tau/\tau_0)^n)$] fits, with $\tau=\tau_1+\tau_2$ (cf.~Fig.~\ref{fig:EchoDecay}), the decay constant $\tau_0$, and an exponent $n$.

The exponents extracted from these fits scatter from 0.5 to 1.3 for the P$_\text{e}$ and P$_\text{n}$ transitions and from 0.5 to 2.6 for the P$_\text{b0}$ transition. In order to compare the time constants from the stretched exponential fits with different exponents, we calculate a mean relaxation time, defined as\cite{NRL6_106}
\begin{equation}
<\tau_0>:=\int_0^{\infty}\text{d}t e^{(t/\tau_0)^n}=\Gamma(1/n)\tau_0/n,
\label{eq:Def_EffTimeconst}
\end{equation}
where $\Gamma$ is the gamma function. If we assume that the physical reason for the stretched exponential decays is the distribution of spin-pair distances,\cite{PRL106_187601,PRL108_27602,suckert_electrically_2013} which also results in a distribution of relaxation times, we can interpret this mean relaxation time of the spin ensemble as an effective time constant of a single exponential decay that represents an average over the relaxation times of the individual spin pairs involved.

To further discuss the results of the decoherence measurements at different radio frequencies, we therefore plot in Fig.~\ref{fig:EchoDecay} (b) the effective echo times $T_\text{2,eff}=\Gamma(1/n)\tau_0/n$ as a function of the normalized gyromagnetic ratio $\gamma_{z,\text{eff}}/\gamma=(\partial\omega/\partial B )/\gamma$; $\omega$ denotes the transition frequency of two states in angular frequency units. Note that $\gamma_{x,\text{eff}}/\gamma$ and $\gamma_{z,\text{eff}}/\gamma$ are a measure for the amount to which a magnetic field couples to $\hat{S}_x$ and $\hat{S}_z$, respectively. Longitudinal and transverse relaxations processes are also described in terms of the $\hat{S}_x$ and $\hat{S}_z$ operators, respectively. If the spin decoherence is caused by magnetic field fluctuations, we expect an increasing spin coherence time with decreasing $\gamma_{z,\text{eff}}/\gamma$. Such magnetic field noise could for example be caused by random spin flips of interface defects coupled to two-level systems in the SiO$_2$.\cite{PRB76_245306} 

As already evident in Fig.~\ref{fig:OverviewLF} (a), the nuclear-spin-like transition frequencies only weakly depend on the external magnetic field reflected in a smaller $\gamma_{z,\text{eff}}/\gamma$, whereas the electron-spin-like transition frequencies exhibit a stronger magnetic field dependence. Considering Fig.~\ref{fig:EchoDecay} (b), we find that indeed by trend, the measured coherence times increase with decreasing $\gamma_{z,\text{eff}}/\gamma$. Especially the electron-spin-like transitions show a roughly linearly increasing coherence time with decreasing $\gamma_{z,\text{eff}}/\gamma$, while the coherence of the nuclear-spin like states does not systematically depend on $\gamma_{z,\text{eff}}/\gamma$. In the following section, we will argue that the weak $\gamma_{z,\text{eff}}/\gamma$ dependence of the coherence time is due to the spin-pair recombination based detection mechanism of pulsed EDMR. We note that the electron spin coherence time of Si:Bi measured by conventional pulsed ESR at $X$-band frequencies systematically depend on the $\partial\omega/\partial B$ value of the considered electron spin transition.\cite{PRB86_245301}

In contrast to the phosphorus spins, the dangling-bond spin exhibits a pronounced (roughly $\propto 1/B$) magnetic field dependence of its coherence time as shown in Fig.~\ref{fig:EchoDecay} (c). We can exclude that the coherence is limited by antiparallel spin-pair recombination, since the corresponding recombination times are of the order of tens of microseconds, as we will see in the following section, significantly longer than the $T_{2\text{,eff}}$ observed. To our knowledge, the microscopic processes that lead to dangling-bond spin decoherence have not been investigated yet. For nitrogen-vacancy (NV) centers in diamond, it has been reported that a nuclear spin bath coupled to the NV electron spin can lead to an artificially shortened coherence time in a similar magnetic field region with the observed coherence time decreasing as a function of $B$.\cite{PRB85_115303} In Sec.~\ref{sec:ESEEM_Pb}, we will present stimulated echo decay measurements of the P$_\text{b0}$ and continue this discussion.

\subsection{Inversion Recovery Measurements: Antiparallel Spin Pair Life Times}\label{sec:InvRec}

\begin{figure*}[!htp]
\includegraphics[]{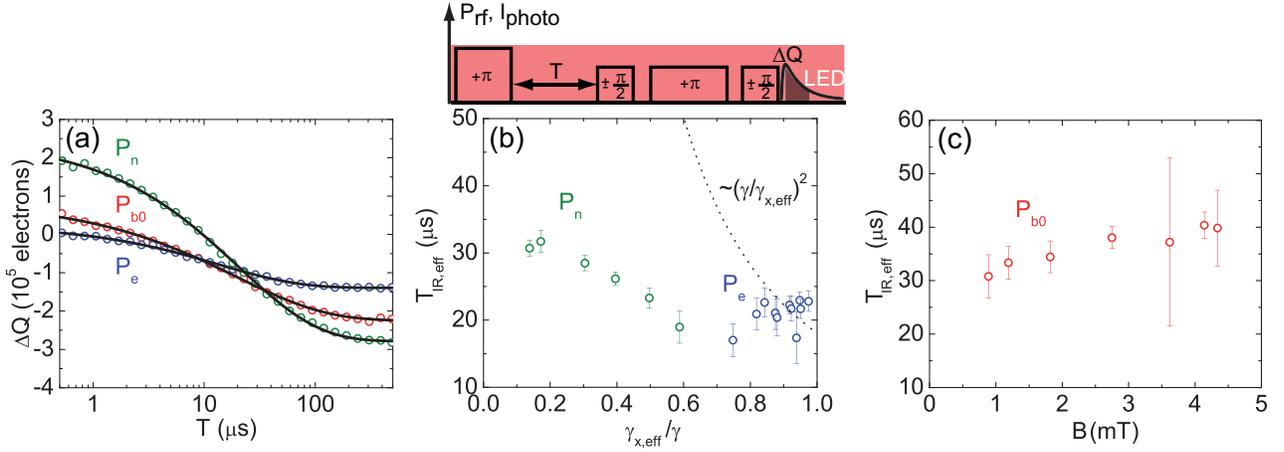}
\caption{On top: Inversion recovery pulse sequence used for the measurements of $\tau_\text{ap}$ in the low-field region. The $\pi$ pulse used for the inversion of the spin system has a larger excitation bandwidth than the $\pi$ pulse in the echo for reasons discussed in the text. (a) Measurements with the pulse sequence sketched on top performed on the P$_\text{b0}$, P$_\text{e}$, and P$_\text{n}$ transitions at an rf of 50.5~MHz with stretched exponential fits. (b) Effective recombination times $T_\text{IR,eff}$ of the P$_\text{e}$ and P$_\text{n}$ transitions as a function of $\gamma_\text{x,eff}/\gamma$. The dashed line shows the theoretically expected $(\gamma/\gamma_\text{x,eff})^2$ dependence of the recombination time; the discrepancy between theory and experiment is discussed in the text. (c) Effective recombination time of the P$_\text{b0}$ transition as a function of the magnetic field.}
\label{fig:InvRec}
\end{figure*}

In this section, we will discuss the life time of antiparallel spin pairs $\tau_\text{ap}$ in the low magnetic field region. We performed inversion recovery measurements\cite{Schweiger_Book,PRB81_75214,hoehne_time_2013} under cw illumination employing the pulse sequence shown in Fig.~\ref{fig:InvRec}. The inversion pulse was applied through a separate pulse channel such that its amplitude could be adjusted independently from the power level used for the detection echo. Typically, the $\pi$-pulse time of the inversion pulses used was a factor of 1.5-2 shorter than the ones of the detection echo, resulting in a larger excitation bandwidth of the inversion pulse compared with the detection echo. Thereby, we achieve that all spins detected with the echo have been inverted. This approach is commonly used for inversion recovery measurements of spin systems where the excitation band width is not large enough to excite the entire resonance line.\cite{Schweiger_Book} The inter-pulse delay of the detection echo was 500~ns for all measurements presented from now on. Figure \ref{fig:InvRec} (a) shows exemplarily the detection echo amplitude as a function of the waiting time $T$ between the inversion pulse and the echo for the P$_\text{b0}$, the P$_\text{e}$, and the P$_\text{n}$ transitions at an rf of 50.5~MHz with stretched exponential fits. The exponents scatter between 0.5 and 0.8. We again calculate effective time constants $T_\text{IR,eff}$ according to Eq.~\eqref{eq:Def_EffTimeconst} as a measure of $\tau_\text{ap}$ and plot the results as a function of $\gamma_\text{x,eff}/\gamma$ for the P$_\text{e}$ and P$_\text{n}$ transitions in Fig.~\ref{fig:InvRec} (b) and as a function of the external magnetic field for the P$_\text{b0}$ transition in Fig.~\ref{fig:InvRec} (c). We note that for an accurate modeling of inversion recovery experiments under cw illumination the illumination-dependent spin-pair generation rate as well as the antiparallel spin pair life time have to be included.\cite{hoehne_time_2013} Fitting inversion recovery data under cw illumination with a single stretched exponential therefore can lead to a systematic error in the determination of the time constant $\tau_\text{ap}$. Since we have performed all experiments under the same illumination intensity, however, the changes in the time constants determined from the inversion recovery decays can still be attributed to a change of $\tau_\text{ap}$.

To justify why we expect the time constants extracted from the inversion recovery measurements to depend on $\gamma_\text{x,eff}/\gamma$ for the $^{31}$P transitions, we make the following consideration. We assume that an rf $\pi$ pulse has brought a $^{31}$P from the $\left|1\right>=\left|\uparrow \Uparrow\right>$ state into the $\left|4\right>=\cos(\eta/2)\left|\downarrow \Uparrow\right>-\sin(\eta/2)\left|\uparrow \Downarrow\right>$ state, cf.~Eq.~\eqref{eq:Eigenstates}, and furthermore suppose that the P$_\text{b0}$ that forms a spin pair with the considered $^{31}$P is in the ``spin up'' state. We project the considered spin pair into a state where the $^{31}$P electron spin and the P$_\text{b0}$ spin are antiparallel, resulting in a matrix element of $\cos(\eta/2)$. Since the recombination rate is proportional to the square of this matrix element,\cite{SSC35_505,PRB80_205206} we expect the recombination rates of the antiparallel spin pairs to be proportional to $(\gamma_\text{x,eff}/\gamma)^2=\cos^2(\eta/2)$ and $(\gamma_\text{x,eff}/\gamma)^2=\sin^2(\eta/2)$ [indicated by the dashed line in Fig.~\ref{fig:InvRec} (b)] for the electron- and nuclear-spin-like transitions, respectively. Note that in this consideration we have assumed the parallel spin pair recombination rate to be zero, which is a good approximation, cf.~the following section.

Considering the experimental results in Fig.~\ref{fig:InvRec} (b), we however observe a much weaker dependence of the inversion recovery times on $\gamma_\text{x,eff}/\gamma$. The experimentally determined time constants vary by roughly a factor of 2 over the considered range of $\gamma_\text{x,eff}/\gamma$, while we would theoretically expect a change by more than a factor of 100, as schematically indicated by the dashed line, extrapolating from the $T_\text{IR,eff}$ values observed at $\gamma_\text{x,eff}/\gamma \approx 0.9 $. This is most likely caused by the different spin pairs with a distribution of P$_\text{b0}$-$^{31}$P distances which exist in the sample. Therefore, we have to expect a broad distribution of recombination rates (assuming an exponential dependence of the recombination rate on the P$_\text{b0}$-$^{31}$P distances).\cite{suckert_electrically_2013} This distribution is manifested, e.g., in the stretched exponential character of the inversion recovery decay as observed in the experiment. A pulsed EDMR experiment is generally only sensitive to a certain window of recombination rates since the recombination time also influences the decay time of the current transient.\cite{hoehne_time_2013} This time window is determined by the box-car integration window and the bandwidth of the transimpedance amplifier. Consequently, spin pairs with faster recombination rates can exist although they are not observed by our present pulsed EDMR experiments.

Reducing the recombination time of all spin pairs by a factor of $(\gamma/\gamma_\text{x,eff})^2$ results in a different subensemble of spin pairs being observed. Experimentally, however, the observed recombination time does not change significantly, as it is mostly determined by the time window. Therefore only a weak dependence of the measured recombination time $T_\text{IR,eff}$ on $\gamma_\text{x,eff}/\gamma$ is observed. Along the same line of argument, it can be understood why the measured  electron- and nuclear-spin coherence times in Sec.~\ref{sec:HahnEcho} only weakly depend on $\gamma_{z,\text{eff}}/\gamma$, assuming that the spin coherence is related to the spin pair distance.

To exclude a strong influence of the generation rate on the mixing angle dependence of $\tau_\text{ap}$ extracted from the inversion recovery experiments under cw illumination, we also performed an inversion recovery experiment under pulsed photoexcitation as introduced in Ref.~\onlinecite{PRL108_27602}. We chose an rf of 77.2~MHz, resulting in $\sin^2(\eta/2)=0.09$ and $\cos^2(\eta/2)=0.85$, for the P$_\text{n}$ and P$_\text{e}$ transitions, respectively. From stretched exponential fits to the data, we found effective time constants of 47~$\mu$s and 30~$\mu$s, for the P$_\text{n}$ and P$_\text{e}$ transitions, respectively, similar to the 47~$\mu$s and 22~$\mu$s extracted from the data under cw illumination, cf.~Fig.~\ref{fig:InvRec}. 
In particular, the theoretically expected $(\gamma_\text{x,eff}/\gamma)^2$ dependence of the recombination rate which would predict roughly a factor of 10 slower time constant for the nuclear-spin like transition is neither observed in the inversion recovery experiments under pulsed photoexcitation nor under cw illumination. Since the signal-to-noise ratio was worse under pulsed photoexcitation, we chose to measure the mixing-angle dependence systematically under cw illumination.

As evident from Fig.~\ref{fig:InvRec} (c), the antiparallel spin-pair recombination time $\tau_\text{ap}$ measured on the P$_\text{b0}$ transition remains approximately constant over the investigated magnetic field range. Comparing the P$_\text{b0}$ recombination times with the P$_\text{e}$ recombination times we note that they differ by up to a factor of 2 [Figs.~\ref{fig:InvRec} (b) and (c)], which seems to be contradicting the assumption of a spin pair. The reason for this is the use of a detection echo with a rather large inter-pulse spacing of 500~ns (to avoid overlapping of the rf pulses which have rather long ring-down times). Since the P$_\text{b0}$ echo decays very fast (cf.~Fig.~\ref{fig:EchoDecay}), we thus select spin-pairs with rather long decoherence times, i.e.~spin pairs which are further apart and therefore recombine more slowly, assuming that a correlation exists between coherence and recombination times of the P$_\text{b0}$. We confirmed this hypothesis experimentally by measuring an inversion recovery with different inter-pulse delays of the detection echo, observing that for short inter-pulse delays the P$_\text{e}$ and P$_\text{b0}$ transitions indeed exhibit the same recombination times (data not shown).

\subsection{Light Pulse Delay Measurements: Parallel Spin Pair Life Times}\label{sec:DPS}
\begin{figure*}[!htp]
\includegraphics[]{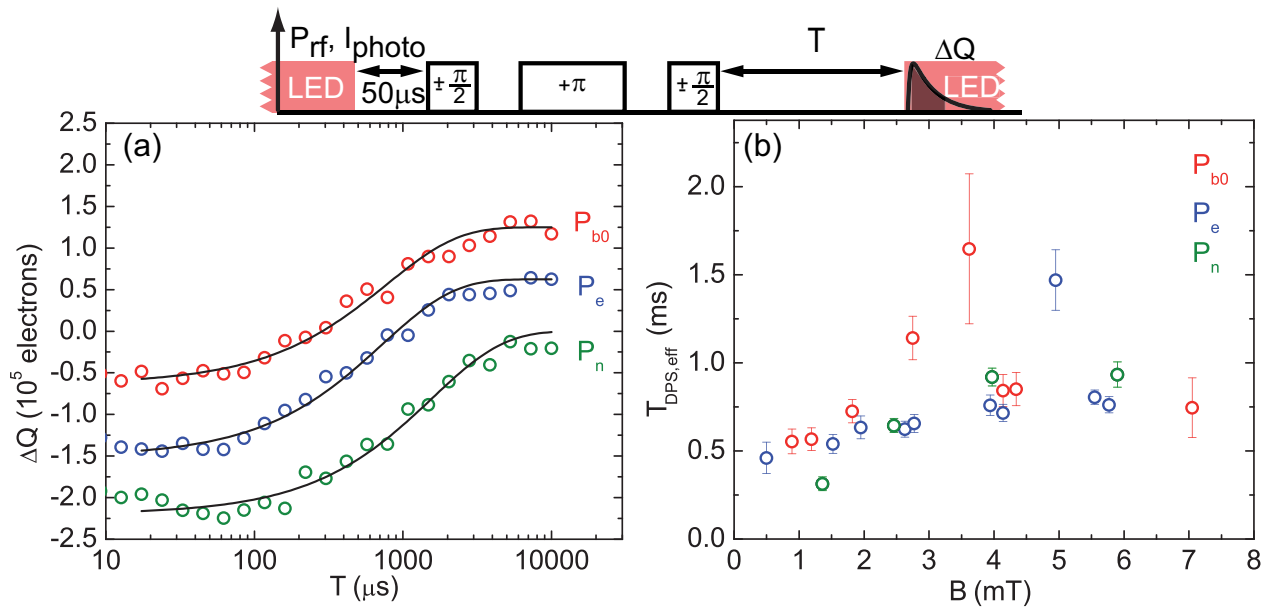}
\caption{On top: Pulse sequence used to determine the lifetime of parallel spin pairs. (a) Measurement of the lifetime of parallel spin pairs $\tau_\text{p}$ for the P$_\text{b0}$, P$_\text{e}$, and P$_\text{n}$ transition at an rf of 50.5~MHz with stretched exponential fits. The data sets are offset for better visibility. (b) The parallel spin pair recombination time as a function of the magnetic field; the exponent was fixed to $n=0.9$.}
\label{fig:DPS}
\end{figure*}

Having discussed the magnetic field dependence of the antiparallel spin-pair recombination time, we turn to the parallel spin-pair lifetime $\tau_\text{p}$. We employ the pulse sequence depicted in Fig.~\ref{fig:DPS}, discussed already in detail in Refs.~\onlinecite{PRL108_27602,hoehne_time_2013}. In Fig.~\ref{fig:DPS} (a), we plot the echo amplitude as a function of the waiting time $T$ for the P$_\text{b0}$, the P$_\text{e}$ and the P$_\text{n}$ transitions, revealing decay time constants of the order of one millisecond. All data were fitted with stretched exponentials with an exponent of $n=0.9$. In Fig.~\ref{fig:DPS} (b) the effective time constants $T_\text{DPS,eff}$  as a measure of $\tau_\text{p}$ [cf.~Eq.~\eqref{eq:Def_EffTimeconst}] are plotted as a function of the external field (DPS stands for ``detection pulse spacing'' since the spacing between the echo and the light pulse during which the photocurrent is detected is varied, cf. Fig.~\ref{fig:DPS}).

If at all, the time constants $T_\text{DPS,eff}$ slightly increase with increasing magnetic field. At $X$-band frequencies, we observed $\tau_\text{p}\approx 2$~ms with an exponent of $n=0.5$,\cite{PRL108_27602} corresponding to an effective time constant of 4~ms according to Eq.~\eqref{eq:Def_EffTimeconst}.
Therefore, it would be interesting to extend the broadband EDMR approach presented in Refs.~\onlinecite{PRL106_37601,RoSI82_74707,N489_541} to pulsed EDMR, in order to systematically study the $B$ dependence of $\tau_\text{p}$ in a larger magnetic field range.

Most likely, the parallel spin pair life time $\tau_\text{p}$ is limited by a spin flip of the P$_\text{b0}$ spin or the $^{31}$P electron spin converting the spin pair to a short-lived antiparallel spin pair that readily recombines. The corresponding longitudinal $^{31}$P electron spin relaxation time $T_1$ is expected to be of the order of seconds under our experimental conditions\cite{PR114_1245} and is thus not expected to have an influence on the parallel spin pair recombination process.
Therefore, we assume $\tau_\text{p}$ to be determined by the dangling-bond spin flip time. While to our knowledge no measurements of the dangling-bond $T_1$ at the crystalline Si/SiO$_2$ interface have been reported, $T_1$ measurements of the dangling bond in amorphous Si and Ge suggest that the $T_1$ relaxation process involves localized two-level states and a phonon,\cite{PRB28_6256,PRB33_4455} resulting in a quadratic temperature dependence. Furthermore, measurements of the dangling-bond $T_1$ of amorphous silicon at 9.3~GHz and 16.5~GHz reveal no dependence of $T_1$ on the magnetic field,\cite{PRB33_4455} consistent with the weak $B$ dependence of $\tau_\text{p}$ observed here. 

Given these results, it seems necessary to measure $\tau_\text{p}$ also as a function of temperature, which is however challenging with the EDMR mechanism employed since the signal typically vanishes at $\gtrsim 10$~K due to the thermal ionization of the P donor. At $X$-band frequencies, we measured $\tau_\text{p}$ in a temperature range of $\approx$~5-10~K and found its value to be temperature independent.

Finally we note that the ratio of $\tau_\text{p}/\tau_\text{ap} \approx 50$ for low magnetic fields, which is comparable to the value obtained at 0.35~T (cf.~Ref.~\onlinecite{PRL108_27602, hoehne_time_2013}). This difference in the time constants is crucial for EDMR experiments because it ensures that in the steady state nearly all spin pairs are in parallel configuration.

\subsection{Stimulated Echo Decay of Dangling Bonds: Low-Field Electron Spin Echo Envelope Modulation}\label{sec:ESEEM_Pb}

As we have seen in Sec.~\ref{sec:HahnEcho}, the Hahn echo decay constant of the dangling-bond transition is rather short and exhibits a $1/B$ dependence. In this section, we further investigate this effect by measuring stimulated echo decays of the P$_\text{b0}$ transition as a function of the rf.
In conventional ESR, the time constant of a stimulated echo decay is given by the longitudinal relaxation time (spin life time) $T_1$ whereas an echo decay is limited by the usually shorter transverse relaxation time (spin coherence time) $T_2$. Therefore, when investigating ESEEM effects, stimulated echo decay measurements allow for a higher spectral resolution than Hahn echo decays, since more oscillation periods can be detected within the echo decay. In addition, slow modulations with periods larger than $T_2$ cannot be observed in Hahn echo while such modulations can be observed in a stimulated echo decay since its decay constant is the usually much longer $T_1$. \cite{Schweiger_Book}

\begin{figure*}[!htp]
\includegraphics[]{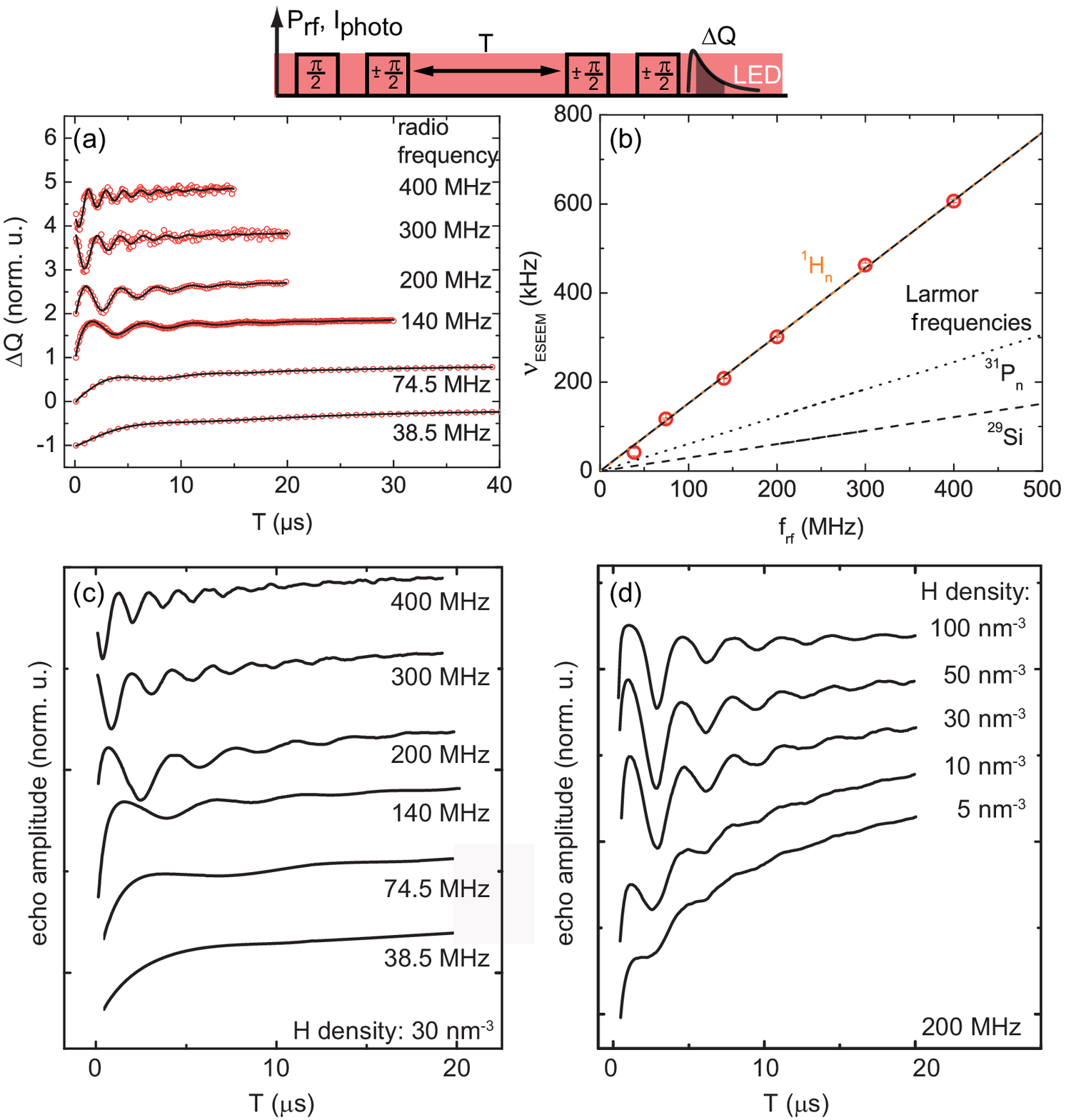}
\caption{On top: Pulse sequence used in the electrically detected stimulated echo decay measurements. (a) Stimulated echo decays of the P$_\text{b0}$ transition measured at different radio frequencies, revealing a pronounced modulation of the echo envelope with a period $T_\text{ESEEM}$. (b) ESEEM frequency $\nu_\text{ESEEM}$ as obtained from fits of the data in (a) plotted against the Zeeman splitting, i.e.~the employed radio frequency $f_\text{rf}$, revealing a linear dependence; the solid black line represents a linear fit to the data through the origin. The free nuclear Larmor frequencies of $^{31}$P (dotted black line), $^{29}$Si (dashed black line), and $^1$H (dashed orange line) are shown for comparison. (c) and (d) The ESEEM signal is simulated based on a point-dipole approximation assuming randomly located H atoms in a 1~nm-thick slab. The plots show the resulting echo signal in normalised units at the same resonance frequencies used in the experiment (c) and for different densities of H atoms (d). To avoid overlapping, the curves in (c) and (d) are vertically offset and the data points at $T=0$~$\mu$s are not shown.}
\label{fig:ESEEM_Pb}
\end{figure*}

In the EDMR experiments presented here, a stimulated spin echo decay is measured by the pulse sequence shown in Fig.~\ref{fig:ESEEM_Pb}. If the two first $\pi/2$ pulses form an effective $\pi$ pulse, the spin pair is in an antiparallel state. Therefore, the signal decays during the waiting period $T$ with a time constant of $\tau_\text{ap}$, rather than $T_1$ as in conventional ESR. As we have seen in the preceding sections, this time constant is still substantially longer than the coherence time of the P$_\text{b0}$. Therefore, it is advantageous to use the stimulated spin echo decay to investigate the possibility of spin echo envelope modulations with relatively small modulation frequencies compared with the coherence time.

The delay between the first and second pulse as well as the third and fourth pulse was set to $\tau_1=200$~ns (cf.~Fig.~\ref{fig:ESEEM_Pb}) and a four-step phase cycle was used to remove all undesired echoes,\cite{Schweiger_Book} to remove the non-resonant background and to realize the lock-in detection scheme discussed in Sec.~\ref{sec:exp_details}. The traces with open circles in Fig.~\ref{fig:ESEEM_Pb} (a) are the resulting stimulated echo decays measured at the indicated rf. The decays have been normalized to an amplitude of one and shifted vertically for better visibility. The black solid lines represent fits using the phenomenological formula
\begin{eqnarray}
\Delta Q_\text{ESEEM,fit}(T)=\Delta Q_0
e^{-(T/\tau_\text{0})^n}\label{eq:LF_ESEEM} \\\nonumber
\times \left\{1 - P e^{-T/\kappa}\left[1-\cos\left(2\pi \nu_\text{ESEEM}T+\phi\right)\right] \right\},
\end{eqnarray}
where $\tau_0$ describes the time constant with which the stimulated echo decays, i.e., the antiparallel spin pair life time $\tau_\text{ap}$, $n$ is an exponent, $P$ is a parameter related to the ESEEM modulation depth,\cite{Schweiger_Book} $\kappa$ is the time constant with which the modulation decays, $\nu_\text{ESEEM}$ is the modulation frequency, $\phi$ a phase shift, and $\Delta Q_0$ an amplitude.

From the fits we extract $\nu_\text{ESEEM}$ and plot it against the Zeeman splitting, i.e., the corresponding rf in Fig.~\ref{fig:ESEEM_Pb} (b), revealing a linear dependence. This is consistent with the fact that the Hahn echo decay constants of the P$_\text{b0}$ transitions shown in Sec.~\ref{sec:HahnEcho} scale with the inverse of the magnetic field, if we assume that the decays shown there are dominated by an echo envelope modulation effect. We further observe that the modulation effect becomes more pronounced with increasing Zeeman frequency and reaches nearly 100\% for an rf of 200~MHz. Extrapolating the ESEEM frequency observed to $X$-band Zeeman splittings would result in an ESEEM frequency of about 15~MHz, which has not been observed in $X$-band ESEEM measurements of comparable samples.\cite{PRL106_196101} At frequencies lower than 38.5~MHz, a modulation of the stimulated echo decay was not observed in our experiments.

To compare the experimental data with the Larmor frequencies of different nuclei possibly involved, we plot the free nuclear spin Larmor frequencies of $^{31}$P, $^{29}$Si, and $^1$H as a function of the electron Zeeman splitting in Fig.~\ref{fig:ESEEM_Pb} (b). From the  the linear fit through the origin we extract a nuclear $g$-factor of 5.58(5) from our ESEEM experiments in agreement with the literature value for hydrogen 5.58569468(6),\cite{ADaNDT90_75} suggesting that the ESEEM effect stems from hydrogen that is weakly hyperfine coupled to the P$_\text{b0}$ electron spin. Since the oxide on the sample is a natural SiO$_2$, it would not be surprising to find hydrogen at the Si/SiO$_2$ interface due to H$_2$O in the natural ambient during the growth of the oxide.\cite{JES126_122} Furthermore, since the sample was neither annealed nore measured in ultra-high vacuum, it will have a thin layer water absorbed on top.\cite{asay_evolution_2005}

The observation of an ESEEM signal exactly at the H nuclear Larmor frequency $\nu_\mathrm{H}$ over a wide range of magnetic fields suggests that the signal originates from a large number of H nuclear spins with small hyperfine interactions compared to $\nu_\mathrm{H}$ (matrix nuclei).\cite{Tsvetkov_Book} From the large modulation depth observed in Fig.~\ref{fig:ESEEM_Pb}(a), we can estimate the density of H atoms. To this end, we calculate the hyperfine interaction between the P$_\mathrm{b0}$ and the H nuclear spins using the point-dipole approximation.\cite{Schweiger_Book}~We assume, that the H atoms are randomly distributed in a 1~nm thick slab corresponding to the thickness of a native oxide\cite{JoAP68_1272}  or a thin water layer\cite{asay_evolution_2005} with an area of $10\times10$~nm$^2$ above the P$_\mathrm{b0}$, similar to the geometry in Refs.~\onlinecite{staudacher_nuclear_2013,mamin_nanoscale_2013}. We further average the resulting ESEEM signal over 400 random configurations of H atoms. In Figs.~\ref{fig:ESEEM_Pb}(c) and (d), we plot the simulated ESEEM signal superimposed on a stretched exponential decay (cf.~Eq.~\eqref{eq:LF_ESEEM}) for different magnetic fields corresponding to the resonance frequencies indicated in  Fig.~\ref{fig:ESEEM_Pb}(c) and for different densities of H atoms  Fig.~\ref{fig:ESEEM_Pb}(d). The results show, that the resonance frequency dependence and the large experimentally observed modulation depth are reproduced well for densities of $\approx$20-50~nm$^{-3}$, which is very near to the number of hydrogen atoms of 66 in a nm$^3$ of water.

These results demonstrate that pulsed low-field EDMR is capable of detecting nuclear magnetic moments in chemi- or physisorbed layers on top of the semiconductor layer where spin-dependent transport is taking place, as it is also observed in ODMR.\cite{staudacher_nuclear_2013,mamin_nanoscale_2013} The high signal-to-noise ratio suggests that
 in EDMR under ultra-high vacuum conditions,  the smaller concentration of H in the oxide or even at the Si/SiO$_2$ interface might be observable,\cite{gerstmann_ab_2010} suggesting a systematic study of hydrogen in differently prepared oxides and under deuterium substitution. This opens up pulsed low-field EDMR as an interesting tool for interface science, ultimately allowing the simultaneous study of the microscopic and electronic structure of surface defects and passivations by magnetic resonance techniques.

\section{Summary and Outlook}\label{sec:Summary_Outlook}

We have demonstrated that the pulsed EDMR methods established in $X$-band can be transferred to the MHz regime, enabling various pulsed low-field experiments in a magnetic field region where the electron Zeeman interaction is comparable to the hyperfine interaction of the $^{31}$P donor. Coherent spin oscillation measurements in this field region revealed a mixing-angle dependence of the Rabi frequency consistent with the crossover from electron- to nuclear-spin-like states and vice versa. The ability to tune the extent to which a state is electron- or nuclear-spin like could be advantageous for realizing a hybrid quantum bit where the state remains nuclear-spin like to store quantum information and is tuned by a fast magnetic field ramp (faster than the decoherence process) to an electron-spin-like state where it can be manipulated on a faster time scale, similar as proposed for Si:Bi at larger magnetic fields.\cite{PRL105_67602} We have investigated the effective coherence times and the effective recombination times of antiparallel and parallel spin pairs in the low field region and have found that we do not observe a strong variation in pulsed EDMR experiments. For the coherence times and the antiparallel spin pair recombination times, we suggest that the reason for this is the detection mechanism, which selects a subensemble of spin-pairs with easily observed recombination and coherence times. The fact that the parallel spin pair recombination time does not show a significant magnetic field trend and that its value is close to the value found in the $X$-band suggests that the dangling-bond spin-flip time $T_1$, which is most likely the time constant limiting the parallel spin-pair lifetime, is nearly magnetic field independent for $B<0.35$~T. 

We have measured Hahn echo and stimulated echo decays of the P$_\text{b0}$ interface state as a function of the magnetic field. We found an effective $1/B$ dependence in the Hahn echo decay which is consistent with the pronounced modulation of the stimulated spin echo decay, which showed a linear $B$ dependence of the ESEEM frequency. The modulation effect is barely visible for an rf of 38.5~MHz, is close to 100\% at 200~MHz and slightly decreases again for higher frequencies, demonstrating the benefit of using low-field pulsed EDMR for the detection of small hyperfine interactions, i.e., hyperfine interactions that are comparable to the nuclear Zeeman interaction. We have found that the observed ESEEM effect can be reproduced in simulations by assuming randomly distributed H nuclei at a density of 20-50~nm$^{-3}$ in a 1~nm thick slab of nuclei above the Si sample. Finally, the feasibility of pulsed low-field experiments demonstrated for the Si:P model system motivates the investigation of other materials with small hyperfine interactions, such as polymers or organic semiconductors for photovoltaics, exploiting the sensitivity of pulsed low-field ESEEM to small hyperfine interactions and the large ENDOR enhancement at low fields.
\begin{acknowledgments}
This work was supported by the Deutsche Forschungsgemeinschaft (Grant No.~Br 1585/5, Br 1585/8, and SFB 631 C3) and by the JST-DFG Strategic Cooperative Program on Nanoelectronics. The work at Keio was supported in part by the Grant-in-Aid for Scientific Research by MEXT, 
in part by NanoQuine, and in part by JSPS Core-to-Core Program. 
\end{acknowledgments}

\end{document}